%% file: FAUST.tex
\documentclass{llncs}

\usepackage{amsmath}
\usepackage{amssymb}
\usepackage{mathrsfs}
\usepackage{graphicx}
\usepackage{color}
\usepackage{algorithm}
\usepackage{algorithmic}

\usepackage{enumitem}
\usepackage{rotating}
\usepackage{tikz}
\newcommand*\circled[1]{\tikz[baseline=(char.base)]{\node[shape=circle,draw,inner sep=0.5pt] (char) {#1};}}
\newcommand*\circledb[1]{\tikz[baseline=(char.base)]{\node[shape=circle,draw,inner sep=0.2pt] (char) {#1};}}

\usepackage{xspace}
\newcommand{\software}{\textsf{FAUST}$^{\mathbbm 2}$\xspace}
\newcommand{\boxname}[1]{\textsf{#1}}

\usepackage{bbm}
\usepackage{pgfplots}
\usepackage{todonotes}
\usepackage{lscape}

\begin{document}

\title{\software: \underline{F}ormal \underline{A}bstractions of \underline{U}ncountable-\underline{ST}ate \underline{ST}ochastic processes}
\author{Sadegh Esmaeil Zadeh Soudjani\inst{1} 
\and Caspar Gevaerts\inst{1}
\and Alessandro Abate\inst{2,1}
}
\authorrunning{Soudjani, Gevaerts, Abate}
\institute{Delft Center for Systems and Control, 
TU Delft -- Delft University of Technology\\
\email{S.EsmaeilZadehSoudjani@tudelft.nl}
\and
Department of Computer Science,  
University of Oxford\\   
\email{alessandro.abate@cs.ox.ac.uk}
}

\maketitle

\begin{abstract}
\software  is a software tool that generates formal abstractions of (possibly non-deterministic) discrete-time Markov processes (dtMP) defined over uncountable (continuous) state spaces. 
A dtMP model (Sec. \ref{dtMPsys}) is specified in MATLAB  
and abstracted as a finite-state Markov chain or Markov decision processes.
The abstraction procedure (Sec. \ref{ABS}) runs in MATLAB and employs parallel computations and fast manipulations based on vector calculus.
The abstract model is formally put in relationship with the concrete dtMP via a user-defined maximum threshold on the approximation error introduced by the abstraction procedure.
\software  allows exporting the abstract model to well-known probabilistic model checkers, such as PRISM or MRMC (Sec. \ref{EI}).
Alternatively, it can handle internally the computation of PCTL properties (e.g. safety or reach-avoid) over the abstract model, and refine the outcomes over the concrete dtMP via a quantified error that depends on the abstraction procedure and the given formula (Sec. \ref{PCTL}). 
The toolbox is available at
\vspace{-0.2cm}
\begin{center}
{\small\texttt{http://sourceforge.net/projects/faust2/}}
\end{center}
\end{abstract}

\section{Models: discrete-time Markov processes}
\label{dtMPsys}
We consider a discrete-time Markov process (dtMP) $s(k), k \in \mathbb N \cup \{0\}$  
defined over a general state space, 
such as a finite-dimensional Euclidean domain \cite{MTH1993} or a hybrid state space \cite{APLS08}. 
The model is denoted by the pair $\mathfrak S = \left(\mathcal S, T_s\right)$. 
$\mathcal S$ is a continuous (uncountable) but bounded state space, e.g. $\mathcal S \subset \mathbb R^n, n < \infty$.
We denote by $\mathcal B (\mathcal S)$ the associated sigma algebra and refer the reader to \cite{APLS08,BS96} for details on measurability and topological considerations.
The conditional stochastic kernel $T_s:\mathcal B(\mathcal S)\times \mathcal S\rightarrow[0,1]$ 
assigns to each point $s \in \mathcal S$ a probability measure $T_s(\cdot | s)$, 
so that for any set $A \in \mathcal B(\mathcal S), k \in \mathbb N \cup \{0\}$,
$\mathbb P(s(k+1) \in A | s(k)=s) = \int_A T_s (dx|s)$.
(Please refer to code or case study for a modelling example.)  

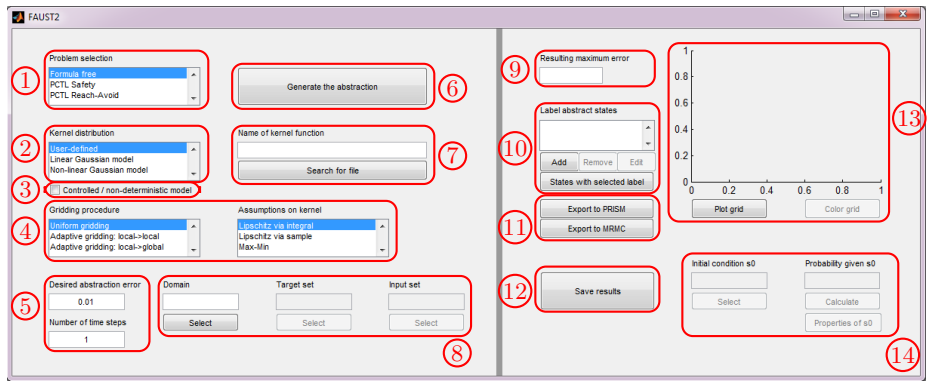
\begin{figure}
\begin{center}
\input{guiwithnumbers.tikz}
\caption{Graphical User Interface of \software, overlaid with numbered boxes}
\label{fig:GUI_Numbered}
\end{center}
\end{figure}

\textit{\textbf{Implementation: }}
The user interaction with \software is enhanced by a Graphical User Interface. 
A dtMP model is fed into \software as follows. 
Select the \boxname{Formula free} option in the box \boxname{Problem selection} \circled{1} in Figure \ref{fig:GUI_Numbered}, 
and enter the bounds on the state space $\mathcal S$ as a $n\times 2$ matrix in the prompt \boxname{Domain} in box \circled{8}. 
Alternatively if the user presses the button \boxname{Select} \circled{8}, a pop-up window prompts the user to enter the lower and upper values of the box-shaped bounds of the state space. 
The transition kernel $T_s$ can be specified by the user (select \boxname{User-defined} \circled{2}) in an m-file, 
entered in the text-box \boxname{Name of kernel function},  
or loaded by pressing the button \boxname{Search for file} \circled{7}. 
Please open the files \texttt{./Templates/SymbolicKernel.m} for a template and \texttt{ExampleKernel.m} for an instance of kernel $T_s$. 
As a special case, 
the class of affine dynamical systems with additive Gaussian noise is described by the difference equation
$s(k+1) = \mathtt{A} s(k)+\mathtt{B}+\eta(k)$, where $\eta(\cdot)\sim\mathcal N(0,\mathtt{Sigma})$. 
(Refer to the Case Study on how to express the difference equation as a stochastic kernel.) 
For this common instance, 
the user can select the option \boxname{Linear Gaussian model} in the box \boxname{Kernel distribution} \circled{2}, 
and input properly-sized matrices \texttt{A,B,Sigma} in the MATLAB workspace.
\software also handles Gaussian dynamical models $s(k+1) = f(s(k))+g(s(k))\eta(k)$ with nonlinear drift and variance:
select the bottom option in box \circled{2} and enter the symbolic function \texttt{[f g]} via box \circled{7}.
\qed

\vspace{-0.2in}

The software also handles models with non-determinism \cite{tmka2013}:
a controlled dtMP is a tuple $\mathfrak S = (\mathcal S,\mathcal U,T_s)$, 
where $\mathcal S$ is as before,
$\mathcal U$ is a continuous control space (e.g. a bounded set in $\mathbb R^m$), 
and $T_s$ is a Borel-measurable stochastic kernel
$T_s:\mathcal B(\mathcal S)\times \mathcal S\times \mathcal U\rightarrow[0,1]$,
which assigns to any state $s\in\mathcal S$ and input $u\in\mathcal U$
a probability measure $T_s(\cdot|s,u)$. 

\textit{\textbf{Implementation: }}
In order to specify a non-deterministic model in \software, 
tick the relevant check \boxname{Controlled/non-deterministic model} \circled{3},  
and enter the bounds on the space $\mathcal U$ as a $m\times 2$ matrix in the window \boxname{Input set} \circled{8}.
\qed

\section{Formal finite-state abstractions of dtMP models}
\label{ABS}

This section discusses the basic procedure to approximate a dtMP $\mathfrak S = (\mathcal S, T_s)$ as a finite-state Markov chain (MC) $\mathfrak P = (\mathcal P, T_p)$, as implemented in \software.  
$\mathcal P = \{z_1, z_2, \ldots, z_p\}$ is a finite set of abstract states of cardinality $p$,  
and $T_p: \mathcal P \times \mathcal P \rightarrow [0,1]$ is a transition probability matrix
over the finite space $\mathcal P$:
$T_p (z, z')$ characterizes the probability of transitioning from state $z$ to state $z'$.  

Algorithm \ref{algo:MC_app} describes the abstraction of model $\mathfrak S$ as a finite-state MC $\mathfrak P$ \cite{APKL10}.  
In Algorithm \ref{algo:MC_app}, 
$\Xi: \mathcal P \rightarrow 2^{\mathcal S}$ represents a set-valued map that associates to any point $z_{i}\in \mathcal P$ the corresponding partition set $A_{i} \subseteq \mathcal S$, 
whereas the map $\xi: 2^\mathcal S \rightarrow \mathcal P$ relates any point $s$ or set in $\mathcal S$ to 
the corresponding discrete state in $\mathcal P$.
\vspace{-0.2in}
\begin{algorithm}[h]
\caption{Abstraction of dtMP $\mathfrak S$ by  MC $\mathfrak P$}
\label{algo:MC_app}
\begin{center}
\begin{algorithmic}[1]
\REQUIRE 
input dtMP $\mathfrak S = (\mathcal S,T_s)$
\STATE
Select a finite
partition of the state space $\mathcal S$ as $\mathcal S = \cup_{i=1}^{p} A_i$  
($A_{i}$ are non-overlapping)
\STATE
For each $A_{i}$, select an arbitrary representative point $z_{i} \in A_{i}, \{z_i\} = \xi (A_i)$ 
\STATE
Define 
$\mathcal P = \{z_i, i=1,...,p\}$ as the finite state space of the MC $\mathfrak P$
\STATE
Compute the transition probability matrix
$T_p (z, z') = T_{s} (\Xi(z')\vert z)$
for all $z,z'\in \mathcal P$
\ENSURE
output MC $\mathfrak P = (\mathcal P, T_p)$
\vspace{-0.1in}
\end{algorithmic}
\end{center}
\end{algorithm}

Consider the representation of the kernel $T_s$ by its density function $t_s:\mathcal S\times\mathcal S\rightarrow \mathbb R^{\ge 0}$, 
namely $T_s(ds'|s) = t_s(s'|s)d s'$ for any $s,s'\in\mathcal S$.
The abstraction error over the next-step probability distribution introduced by Algorithm \ref{algo:MC_app} depends on the regularity of function $t_s$: 
assuming that $t_s$ is Lipschitz continuous, namely that there is a finite positive constant $h_s$ such that
\begin{equation}
\label{eq:glob_lip}
\left|t_{s}(\bar{s}\vert s)-t_{s}(\bar{s}\vert s')\right|\leq h_s\left\|s-s'\right\|, \quad\forall s,s',\bar s \in \mathcal S,
\end{equation}
then the next-step error is $E = h_s\delta_s\mathscr L(\mathcal S)$, 
where $\delta_s$ is the max diameter of the state-space partition sets and $\mathscr L(\mathcal S)$ is the volume of the state space \cite{APKL10}. 
When interested in working over a finite, $N$-step time horizon, the error results in the quantity $E N$. 
Notice that the error can be reduced via $\delta_s$ by considering a finer partition, 
which on the other hand results in a MC $\mathfrak P$ with a larger state space. 

\textit{\textbf{Implementation: }}
\software enables the user to enter the time horizon $N$ of interest (box \boxname{Number of time steps} \circled{5}), 
and a threshold on the maximum allowed error (box \boxname{Desired abstraction error} \circled{5}). 
The software generates a Markov chain with the desired accuracy by pressing the button \boxname{Generate the abstraction} \circled{6}.
Among other messages, 
the user is prompted with an estimated running time, 
which is based on an over-approximation of the Lipschitz constant of the kernel, 
on a uniform partitioning of the space $\mathcal S$ \footnote{ 
At the moment we assume to have selected options \boxname{Uniform gridding} and \boxname{Lipschitz via integral} among the lists in box \circled{4}. 
Comments on further options are in Section \ref{PCTL}. 
}, 
and on the availability of parallelization procedures in MATLAB, 
and is asked whether to proceed.
\qed 

In the case of a non-deterministic dtMP, 
the input space is also partitioned as $\mathcal U = \cup_{i=1}^{q}U_i$, 
and arbitrary points $u_i\in U_i$ are selected. 
The dtMP $\mathfrak S$ is abstracted as a Markov decision process (MDP) 
$\mathfrak P = (\mathcal P, \mathcal U_p, T_p)$, 
where now the finite input space is $\mathcal U_p = \{u_1,u_2,\ldots,u_q\}$, 
and $T_p (u, z, z') = T_{s} (\Xi(z')\vert z,u)$ for all $z,z'\in\mathcal P, u\in\mathcal U_p$.
The abstraction error can be formally quantified as $E = 2(h_s\delta_s+h_u\delta_u)\mathscr L(\mathcal S)$, 
where $\delta_u$ is the max diameter of the input-space partitions and $h_u$ is the Lipschitz constant of the density function with respect to the inputs, 
that is
$\left|t_{s}(\bar{s}\vert s,u)-t_{s}(\bar{s}\vert s,u')\right|\leq h_u\left\|u-u'\right\|$, 
$\forall u,u'\in\mathcal U,s,\bar s \in \mathcal S$.

\textit{\textbf{Implementation: }}
The user may tick the check
in \circled{3} to indicate that the dtMP is controlled (non-deterministic), 
specify a box-shaped domain for the input in box \boxname{Input set} \circled{8},  
enter a time horizon in box \boxname{Number of time steps} \circled{5}, 
and require an error threshold in box \boxname{Desired abstraction error} \circled{5}. 
\software automatically generates an MDP according to the relevant formula on the error.   

Notice that the quantification of the abstraction error requires state and input spaces to be bounded.  
In case of an unbounded state space, 
the user should truncate it to a bounded, box-shaped domain:
selecting the \boxname{Formula free} option in the box \boxname{Problem selection} \circled{1},
the domain is prompted in box \boxname{Domain} \circled{8}.
Algorithm \ref{algo:MC_app} is automatically adjusted by assigning an absorbing abstract state to the truncated part of the state space. 
For details please see \cite{TA13,SA13}. 
\qed 

The states of the abstract model $\mathfrak P$ may be labeled. 
The state labeling map $\mathsf L:\mathcal P\rightarrow \Sigma$, 
where $\Sigma$ is a finite alphabet, 
is defined by a set of linear inequalities:  
for any $\alpha\in\Sigma$ the user characterises the set of states $\mathsf L^{-1}(\alpha)$ as the intersection of half-planes (say, as a box or a simplex):  
the software automatically determines all points $z\in\mathcal P$ belonging to set $\mathsf L^{-1}(\alpha)$.   
The obtained labeled finite-state model
can be automatically exported to well-known model checkers, 
such as PRISM and MRMC \cite{HKNP06,KKZ05}, 
for further analysis.   
In view of the discussed error bounds, 
the outcomes of the model checking procedures over the abstract model $\mathfrak P$ may be refined over the concrete dtMP $\mathfrak S$ -- 
more details can be found in \cite{APKL10,TA13}. 

\textit{\textbf{Implementation: }}
Labels are introduced in \software as follows: 
suppose that the intersection of half-planes $A_{\alpha}z\le B_{\alpha}$ (where $A_{\alpha},B_{\alpha}$ are properly-sized matrices) tags states $z$ by label $\alpha\in\Sigma$. 
The user may add such a label
by pressing button \boxname{Add} \circledb{10} and subsequently entering symbol $\alpha$ and matrices $A_{\alpha},B_{\alpha}$ in the pop-up window.
The user can also edit or remove any previously defined label using buttons \boxname{Edit, Remove} in \circledb{10}, respectively. 
The button \boxname{States with selected label} \circledb{10} shows the sets associated to the active label over the plot in \circledb{13} . 

The user may click the buttons in \circledb{11} to export the abstracted model to PRISM or to MRMC. 
Alternatively, \software is designed to automatically check or optimize over (quantitative, non-nested) PCTL properties, 
without relying on external model checkers: Section \ref{PCTL} elaborates on this capability. 
\qed

\vspace{-0.1in}
\section{Formula-dependent abstractions for verification}
\label{PCTL}

Algorithm \ref{algo:MC_app}, presented in Section \ref{ABS}, 
can be employed to abstract a dtMP as a finite-state MC/MDP, 
and to directly check it against properties such as probabilistic invariance or reach-avoid, 
that is over (quantitative, non-nested) bounded-until specifications in PCTL \cite{HJ94}. 
Next, we detail this procedure for the finite-horizon probabilistic invariance (a.k.a. safety) problem, 
which can be formalized as follows. 
Consider a bounded continuous set $A \in \mathcal B(\mathcal S)$ representing the set of safe states.
Compute the probability that an execution of $\mathfrak S$, 
associated with an initial condition $s_{0} \in \mathcal S$ 
remains within set $A$ during the finite time horizon $[0,N]$, that is
$p_{s_0}(A) := \mathbb P\{ s(k)\in A \text{ for all } k\in [0,N]| s(0) = s_{0} \}$.

The quantity $p_{s_0}(A)$ can be employed to characterise the satisfiability set of a corresponding bounded-until PCTL formula, namely
\vspace{-0.075in}
\begin{equation*}
s_0  \models
\mathbb P_{\sim\epsilon} \{\textsf{true}\,\, \mathsf{U}^{\leq N} (\mathcal S\backslash A) \}
\quad
\Leftrightarrow
\quad
p_{s_0}(A)\backsim 1-\epsilon,
\vspace{-0.075in}
\end{equation*}
where 
$\mathcal S\backslash A$ is the complement of $A$ over $\mathcal S$, 
$\mathsf{true}$ is a state formula valid everywhere on $\mathcal S$, 
the inequality operator $\sim\, \in \{>,\geq, <, \leq\}$, 
and $\backsim$ represents its complement.

\software formally approximates the computation of $p_{s_0}(A), \forall s_0 \in \mathcal S$, as follows.  
$\mathfrak S$ is abstracted as an MC $\mathfrak P$ via Algorithm \ref{algo:MC_app}: 
the bounded safe set  $A$ is partitioned as $A = \cup_{i=1}^{p-1} A_i$; 
representative points $z_i\in A_i$ are selected and, along with an extra absorbing variable $\phi$ for $\mathcal S\backslash A$, make up the state space $\mathcal P$; 
the transition probability matrix $T_p$ is obtained by marginalising the concrete kernel $T_s$. 
Given the obtained discrete-time MC $\mathfrak P = (\mathcal P, T_p)$ and considering the finite safe set $A_p = \{z_1,\ldots,z_{p-1}\} \subset \mathcal P$, 
\software internally computes the safety probability over $\mathfrak P$ via dynamic programming \cite{APKL10}, 
along with the associated abstraction error which is now tailored over the PCTL formula of interest.

\textit{\textbf{Implementation: }}
The user may select option \textsf{PCTL Safety} in the list within box \circled{1}, 
enter the boundaries of the \textsf{Safe set} within box \circled{8}, 
and press button \circled{6} to proceed obtaining the abstraction and computing the probability of the selected formula.
The computed value of $p_{s_{0}}(A)$ is displayed in box \boxname{Probability given s0} \circledb{14}, 
for any user-selected initial state $s_0$ that is input in box \boxname{Initial condition s0} \circledb{14}.
The user can optionally press button \boxname{Properties of s0} \circledb{14} to get more information about the concrete state $s_0$, 
including the related discrete state $z = \xi(\Xi(s))$ of the MC, as well as the associated labels.
Furthermore, 
the quantity $p_{s_{0}}(A)$ can be plotted, 
as a function of the initial state $s_0$, 
by pressing buttons \boxname{Plot grid} and \boxname{Color grid} in \circledb{13}.
Clearly these outputs are exclusively available for models of dimensions $n=1,2,3$.   
\qed

It is of interest to obtain tight bounds on the error associated to the abstraction procedure since, 
given a user-defined error threshold,   
tighter bounds would generate abstract models $\mathfrak P$ with fewer states. 
The abstraction error bound in \software, tailored around the discussed safety problem, 
can be efficiently decreased under different types of regularity assumptions on the conditional density function of the dtMP $\mathfrak S$ \cite{SA13}.  
For instance, in contrast to the global continuity assumption in \eqref{eq:glob_lip}, 
if $t_s$ is locally Lipschitz continuous as 
\begin{equation}
\label{eq:local_lip}
\left|t_{s}(\bar{s}\vert s)-t_{s}(\bar{s}\vert s')\right|\leq h(i,j)\left\|s-s'\right\|, \quad\forall\bar{s}\in A_{j}, \forall s,s'\in A_{i}, 
\end{equation}
(here sets $A_i$ form a partition of $A$, as from Algorithm \ref{algo:MC_app})  
then the error is 
\begin{equation}
\label{eq:boundSHS}
\vert p_{s_{0}}(A) - p_{p_{0}}(A_p)\vert\leq \max\{\gamma_{i}\delta_{i}\vert i=1,...,p\}, 
\end{equation}
where $p_{p_{0}}(A_p)$ is initialized at the discrete state $p_0=\xi(s_0)\in A_p$. 
Here $\delta_{i}$ is the diameter of the set $A_{i}\subset A$,
and the constants $\gamma_{i}$ are given by 
$\gamma_{i} = N\sum_{j=1}^{m}h (i,j) \mathscr L(A_{j})$. 
Since $h(i,j) \leq h_s$, the obtained error in \eqref{eq:boundSHS} is smaller than the older quantity $N h_s\delta_s\mathscr L(\mathcal S)$.
Notice that the structure of the error in \eqref{eq:boundSHS}
leads to gridding algorithms for abstraction that are adapted to the formula and can be made sequential \cite{SA13}: 
\software initialises the procedure with coarse partition sets (resulting in a small MC abstraction but with a large approximation error), 
and sequentially refines the partitions adaptively where the local errors are high 
(leading to an MC abstraction with increasing state space),  
until the global error becomes less than a user-defined threshold. 

\textit{\textbf{Implementation: }}
\software allows the user to select three different gridding procedures in box \boxname{Gridding procedure} \circled{4}:  
the reader is referred to \cite{SA13} for the details of these three options. 
The \boxname{Uniform gridding} option leads to a one-shot (non sequential) procedure, as already discussed in Section \ref{ABS}, 
whereas the two \boxname{Adaptive gridding} options result in sequential and adaptive procedures leading to better errors and to smaller abstractions,   
but in general requiring more computation time. 
The error bound quantification hinges on the constant in the right-hand side of \eqref{eq:local_lip},  
which can be computed differently as in box \boxname{Assumptions on kernel} \circled{4}: tighter errors lead to longer computations \cite{SA13}. 
In order to provide with full control on the chosen inputs, 
for any possible selection of gridding procedure, desired abstraction error, and error bound computation, 
the user is prompted in a pop-up window with an estimated running time, and asked whether to proceed. 

This range of algorithms and procedures are also implemented for probabilistic reach-avoid (constrained reachability) problems, 
which are encompassed by general bounded-until PCTL formulas $\mathbb P_{\sim\epsilon} \{\Phi\,\, \mathsf{U}^{\leq N} \Psi \}$. 
The user can select this option in box \boxname{Problem selection} \circled{1}, 
and is asked to input sets $\Phi, \Psi$ as safe and target sets in the texts in box \circled{8}.

Let us remark that the described abstraction algorithms and procedures are as well available for the formula-free abstraction discussed in Section \ref{ABS}. 
\qed

The safety problem for a controlled dtMP \cite{tmka2013} is here defined as finding the \emph{maximally safe} deterministic Markov policy $\pi^\ast$, 
such that $p_{s_0}^{\pi^\ast}(A) = \sup_{\pi}p_{s_0}^{\pi}(A)$, $\forall s_0\in A$, 
where $p_{s_0}^{\pi}(A)$ is the safety probability under given policy $\pi = (\mu_0,\mu_1,\cdots),$ $\mu_k:\mathcal S\rightarrow \mathcal U$. 
Similarly we can compute the \emph{minimally safe} policy, 
or an optimal policy for the reach-avoid problem.  

\textit{\textbf{Implementation: }}
\software computes a suboptimal policy for a given problem over an MDP, 
with a given threshold on the distance to the optimal safety probability, 
and quantifies the corresponding approximate quantity $p_{s_0}^{\pi^\ast}(A)$.
The approximate optimal policy can be stored by pushing button \boxname{Save results} \circledb{12}, 
which provides the user with two options: 
either storing it in the disk as a .mat file, 
or loading it to the workspace. 
\qed

\vspace{-0.15in}
\section{Extensions and outlook}
\label{EI}

\software is presently implemented in MATLAB, 
which is the modelling software of choice in a number of engineering areas. 
We plan to improve part of its functionalities on a lower-level programming language.
We plan to extend the functionality of \software by allowing for general label-dependent partitioning,
and we are exploring the implementation with otherwise-shaped partitioning sets \cite{SA13}.
We further plan to extend the applicability of \software to models with 
discontinuous and degenerate \cite{SA12,SATAC12} kernels, 
to implement higher-order approximations \cite{SAH12}, 
and to embed formal truncations of the model dynamics \cite{SA14}.
Finally, we plan to look into developing bounds for infinite horizon properties.

\newpage
\bibliographystyle{splncs}

\section*{Appendix} 

\subsection*{Accessing and testing \software}
The toolbox is available at
\vspace{-0.2cm}
\begin{center}
{\small\texttt{http://sourceforge.net/projects/faust2/}}
\end{center}
This toolbox has been successfully tested with MATLAB R2012a, R2012b, R2013a, R2013b, 
on machines running Windows 7, Apple OSX 10.9, and Linux OpenSUSE.  
\software exploits the command \textsf{integral} of MATLAB (introduced in version R2012a) for numerical integrations. 
(The previous versions of MATLAB contain instruction \textsf{quad} and its variations, 
which will be removed in the future versions of MATLAB -- 
we have thus opted for the most up-to-date version.)
Optimization and symbolic computation toolboxes of MATLAB are necessary.
\software automatically checks the presence of these packages and displays an error to the user in their absence.
The software also takes the advantage of the MATLAB parallel computation toolbox if present.     
The use of parallel computation toolbox is currently disabled for Apple operating systems due to a conflict. 

Please download \software from Sourceforge. 
The files are organized in the main folder as follows:
the sub-folder \texttt{Autonomous Models} contains the codes for deterministic systems (without input);
the sub-folder \texttt{Controlled Models} includes the codes for non-deterministic systems (input dependent);
the sub-folder \texttt{Templates} contains templates and examples for the definition of symbolic conditional density functions; 
the sub-folder \texttt{Case Study} contains the files used in the next Section to test the software on a practical study.
The file \texttt{README} can be opened with your preferred text editor and contains instructions on how to set up and run the software.
Alternatively, in order to ease the job of CAV14 reviewers, 
\software can be tested on a case study as elaborated in the next Section.  
Please set the current directory of MATLAB to the folder where the software is stored and run \texttt{FAUST2.m} from the MATLAB command line. 

\subsection*{Case study}
\label{CB}

In this section we apply \software to compute optimal control strategies for the known room temperature regulation benchmark \cite{FI04}.  
Probabilistic models for the underlying dynamics are based on \cite{MC85} and on \cite{APLS08}.
We consider the temperature regulation in multiple rooms via cooling water circulation.
The amount of extracted heat is changed via a flow-control valve.
Then the input signal is the percentage of the valve in the open position. 
The dynamics of the room temperature evolve in discrete time according to the equations 
\begin{align}
s_1(k+1)& = s_1(k)+\frac{\Delta}{C_{ra}}((s_2(k)-s_1(k))k_{cw}u(k)+(T_a-s_1(k))k_{out})+\eta_{ra}(k),\nonumber\\
s_2(k+1)& = s_2(k)+\frac{\Delta}{C_{cw}}((s_1(k)-s_2(k))k_{cw}u(k)+Q)+\eta_{cw}(k),\label{eq:temp_dyn}
\end{align}
where $s_1$ is the air temperature inside the room,
$s_2$ is the cooling water temperature,
$T_a$ is the ambient temperature,
$\Delta$ is the discrete sampling time [min],
and
$\eta_{ra}(\cdot),\eta_{cw}(\cdot)$ are stationary, independent random processes 
with normal distributions $\mathcal{N}(0,\sigma_{ra}^2\Delta)$ and $\mathcal{N}(0,\sigma_{cw}^2\Delta)$, respectively.
Equations \eqref{eq:temp_dyn} can be encompassed in the condensed two-dimensional model 
\begin{equation*}
s(k+1) = f(s(k),u(k))+\eta(k),\quad \eta(\cdot)\sim \mathcal N(0,\varSigma_\eta),
\end{equation*}
which results in a stochastic kernel that is a Gaussian conditional distribution $\mathcal N(f(s,u),\varSigma_\eta)$, 
where $\varSigma_\eta = diag(\Delta[\sigma_{ra}^2,\sigma_{cw}^2])$. 
The file \texttt{Chiller\_Kernel\_2d.m}
appearing with the release of the software,
provides numerical values and physical interpretations of the parameters in equations \eqref{eq:temp_dyn}, 
as well as the symbolic structure of the conditional density function.
The dynamical model in \eqref{eq:temp_dyn} can be as well extended to a two-room temperature control (which results in a three-dimensional model), 
and its conditional density function can be found in file \texttt{Chiller\_Kernel\_3d.m}. 
We will run \software on both 2D and 3D setups. 

We are interested in keeping the temperature of the room(s) within a given temperature interval over a fixed time horizon: 
this can be easily stated as a (probabilistic) safety problem, 
where we maximise over the feasible inputs to the model. 
We instantiate and compute this problem over the model above as described in the main text, 
while providing a step-by-step guide to the user.

In order to select the problem and import the model in \software, please follow these steps:
select \boxname{PCTL Safety} in box \circled{1},
choose \boxname{User-defined} in box \circled{2},
tick the check-box \circled{3} to indicate a controlled model,
and write the name \boxname{Chiller\_Kernel\_2d.m} in the text of box \circled{7} to load the density function of the two-dimensional model \eqref{eq:temp_dyn}.

In the next stage we perform the abstraction and compute the quantity of interest (maximal safety probability). 
Select the most straightforward (but coarsest) abstraction algorithm, 
by choosing options \boxname{Uniform gridding} and \boxname{Lipschitz via integral} in \circled{4}. 
Proceed entering the problem parameters as follows:
input the number of time steps as \texttt{3} and select a desired abstraction error equal to $0.5$ in box \circled{5}; 
enter the safe temperature interval $A$ as \texttt{[19.7,20.3; 4.7,5.3]}, 
as well as the input space $\mathcal U$ as \texttt{[0,1]}
in the text within box \circled{8}. 

\begin{figure}
\begin{center}
\includegraphics[scale = 0.4]{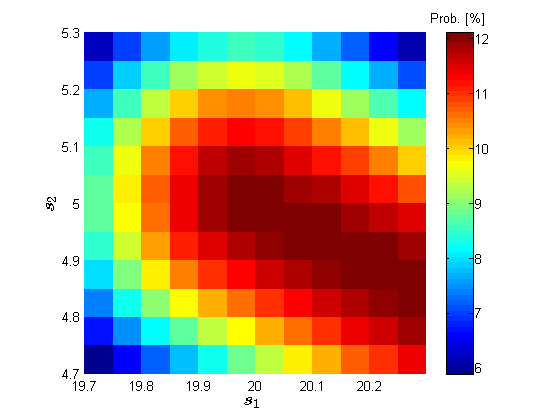}
\includegraphics[scale = 0.4]{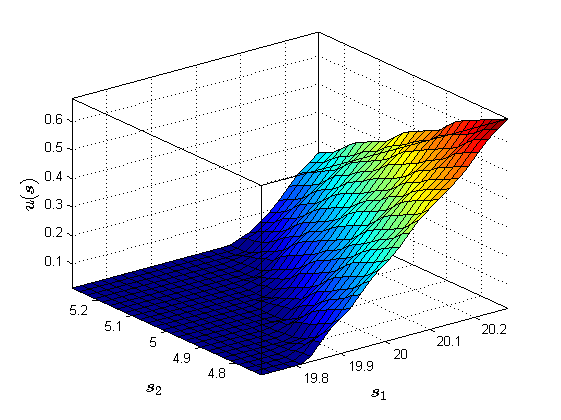}
\caption{Room temperature control problem. 
Left: obtained uniform partition of the safe set, along with optimal safety probability for each partition set (colour bar on the right).  
The safety probability is equal to zero over the complement of the safe set. 
Right: optimal Markov policy at step $N-1$, as a function of the state.}
\label{fig:ss_2d}
\end{center}
\end{figure}

At this point the software can proceed with the main computations. 
Please press the button in box \circled{6}, 
in order to generate the abstract MDP, to compute the optimal policy and the related maximal safety probability. 
When the computation is complete, 
let us proceed with some post-processing: 
press the buttons \boxname{Plot grid} and \boxname{Color grid} in box \circledb{13}, 
to generate Figure \ref{fig:ss_2d} (left) representing the maximal safety probability. 
The result of the computation can be stored for further analysis by pressing button \circledb{12}: 
for instance Figure \ref{fig:ss_2d} (right) is generated by retrieving the optimal state-dependent Markov policy at step $N-1$.

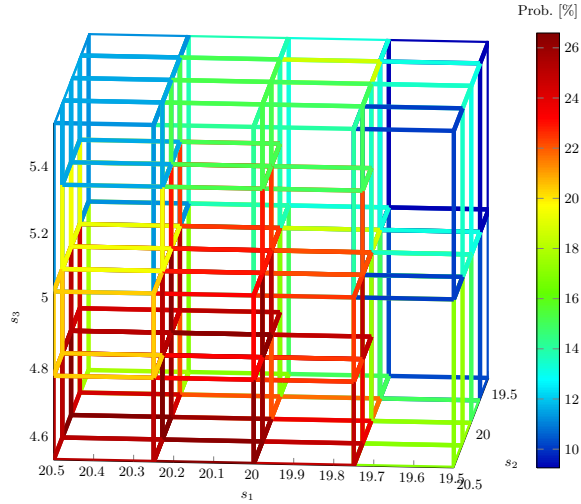
\begin{figure}
\begin{center}
\scalebox{0.6}{\input{states_3d.tikz}}
\caption{Two-room temperature control problem. 
Obtained partition of the safe set, together (bar) with optimal safety probability.}
\label{fig:ss_3d}
\end{center}
\end{figure}

A similar procedure can be followed to study the same probabilistic safety problem over a two-room temperature control, 
instantiated via the density function \texttt{Chiller\_Kernel\_3d.m}.  
Figure \ref{fig:ss_3d} presents the outcomes obtained using the \boxname{Adaptive gridding} and \boxname{Lipschitz via integral} options,
selected in box \circled{4}. The abstraction parameters used in this problem is as follows:
number of time steps \texttt{3},
safe temperature interval \texttt{[19.5,20.5; 19.5,20.5; 4.5,5.5]},
input space \texttt{[0,1; 0,1]}.
We have selected a large abstraction error equal to $12$ in box \circled{5} to be able to visualize the adaptive grid generated by the software. 
The user can choose a smaller error at the cost of a larger computation time.

\subsection*{Summary of the commands in the Graphical User Interface, Figure \ref{fig:GUI_Numbered}}

We provide a summary of the commands of the GUI in \software, 
as they appear in the boxes highlighted in Figure \ref{fig:GUI_Numbered}. 

\begin{itemize}
\item[\circled{1}]
The box \boxname{Problem selection} provides a list with three options: 
select \boxname{Formula free} to obtain an abstraction of the model which can be exported to PRISM or to MRMC for further analysis; 
choose \boxname{PCTL Safety} in order to abstract the model and compute a safety probability; 
or opt for \boxname{PCTL Reach-Avoid} to get the abstraction tailored around the computation of the reach-avoid probability. 

\item[\circled{2}]
The box \boxname{Kernel distribution} gives three options in a list: 
select \boxname{Linear Gaussian model} if the model belongs to the class of Linear Gaussian difference equations (cf. Section \ref{dtMPsys})
and define matrices \texttt{A,B,Sigma} in the MATLAB workspace;
choose \boxname{Non-linear Gaussian model} if the process noise is Gaussian and the drift and variance are non-linear (cf. Section \ref{dtMPsys}),
enter the drift and variance as a single symbolic function with two outputs via box \circled{7};
otherwise choose \boxname{User-defined} and enter your kernel as a symbolic function using \circled{7}.

\item[\circled{3}]
Check this box if the model is non-deterministic (controlled).

\item[\circled{4}]
Box \boxname{Gridding procedure} provides three options:
select \boxname{Uniform gridding} to generate a grid based on global Lipschitz constant $h$ (cf. Section \ref{ABS}), 
where the state space is partitioned uniformly along each dimension; 
choose \boxname{Adaptive gridding: local\texttt{->}local} to generate the grid adaptively based on local Lipschitz constants $h(i,j)$ (cf. Section \ref{PCTL}), 
where the size of partition sets is smaller where the local error is higher; 
select \boxname{Adaptive gridding: local\texttt{->}global} to generate the grid adaptively based on local Lipschitz constants $h(i)$ (cf. \cite{SA13}). 
The first option is likely to generate the largest number of partition sets and to be the fastest in the generation of the grid. 
The second option is likely to generate the smallest number of partition sets but to be the slowest in the grid generation.
For the detailed comparison of these gridding procedures, please see \cite{SA13}. 

The box \boxname{Assumptions on kernel} provides three choices: 
option \boxname{Lipschitz via integral} requires the density function $t_s(\bar s|s)$ to be Lipschitz continuous with respect to the current state $s$, 
and the quantity $T_p (z, z') = T_{s} (\Xi(z')\vert z)$ is used in the marginalisation (integration) step; 
option \boxname{Lipschitz via sample} requires the density function $t_s(\bar s|s)$ to be Lipschitz continuous with respect to both current and the next states $s,\bar s$,
and the quantity $T_p (z, z') = T_{s} (z'\vert z)\mathscr{L}(\Xi(z'))$ is used in the marginalisation step; 
option \boxname{Max-Min} does not require any continuity assumption, but takes longer time in the computation of the error.

\item[\circled{5}]
The time horizon of the desired PCTL formula or of the problem of interest, 
and the required upper bound on the abstraction error should be input in these two boxes. 
For the case of formula-free abstraction you may enter \texttt{1} as the number of time steps. 

\item[\circled{6}]
Press this button after entering the necessary data to generate the abstraction: this runs the main code. 
First, various checks are done to ensure the correctness of the inputed data. 
Then the partition sets are generated via gridding, 
the transition matrix is calculated, 
and the probability and the optimal policy are computed if applicable.    

\item[\circled{7}]
This box is activated for options \boxname{User-defined} and \boxname{Non-linear Gaussian model} in \circled{2}.
For the first option, the conditional density function must be an m-file that generates $t_s(\bar s|s,u)$ symbolically. 
Please refer to \texttt{SymbolicKernel.m} for a template and \texttt{ExampleKernel.m} for an example.
The name of kernel function should be entered in the text-box or the function should be loaded by pressing the button \boxname{Search for file}.
For the option \boxname{Non-linear Gaussian model}, the non-linear drift and variance must be specified as a single symbolic function with two outputs. Please refer to \texttt{NonLinKernel.m} for a template and \texttt{NonLinKernelExample.m} for an example.

\item[\circled{8}]
If the \boxname{Formula-free} option is selected in \circled{1}, 
the user can enter the bounds of the state space in the first of the boxes, named \boxname{Domain}. 
In case any of the additional two options in \circled{1} are selected,
the boundaries of the safe set should be entered in the first text-box named \boxname{Safe set}.
If the \boxname{PCTL Reach-Avoid} option in \circled{1} is selected,
the second box is activated and the boundaries of the target set should be entered in the text-box named \boxname{Target set}.
If the model is non-deterministic and the check in box \circled{3} is ticked, 
the third box is also activated and the boundaries of the Input space may be entered in the box named \boxname{Input set}. 
In all cases the boundaries are to be given as a matrix with two columns, 
where the first and second columns contain lower and upper bounds, respectively.
Alternatively, the user can press the \boxname{Select} button and separately enter the lower and upper bounds in the pop-up window. 

\item[\circled{9}]
The resulting error of the abstraction procedure, 
which is less than or equal to the desired abstraction error introduced in \circled{5}. 
This box shows the error associated to the abstracted model. 

\item[\circledb{10}]
The user can add, remove, or edit labels associated to the abstract states.
The set of states with any label $\alpha\in\Sigma$ can be represented by the intersection of half-planes $A_{\alpha}z\le B_{\alpha}$.
In order to tag these states with the associated label,
the user presses button \boxname{Add} and subsequently enters symbol $\alpha$ and matrices $A_{\alpha},B_{\alpha}$ in a pop-up window.
The user can also edit or remove any previously defined label by activating its symbol in the static-box and using buttons \boxname{Edit, Remove}.
The button \boxname{States with selected label} will show the set of states associated with the active label in \circled{13}.
Adding labels is essential in particular for exporting the result to PRISM or to MRMC. 

\item[\circledb{11}]
The abstracted Markov chain or MDP can be exported to PRISM or to MRMC using these buttons. 
\software enables two ways of exporting the result to PRISM: as a \texttt{.prism} format that is suitable for its GUI, 
or as the combination of \texttt{.tra} and \texttt{.sta} files, 
which are appropriate for the command line.

\item[\circledb{12}]
Use this button to store the results. 
A pop-up window appears after pushing the button and the user can opt for storing the date over the workspace, or in memory as an \texttt{.mat} file.

\item[\circledb{13}]
The user can plot the generated grid for the state space using the first button. 
Pressing this button opens a new window showing the partitioned input space for the controlled model. 
The solution of the safety and of the reach-avoid probability can also be visualized by pressing the second button. 
This option obviously works exclusively for dimensions $n=1,2,3$. 

\item[\circledb{14}]
The user can enter any initial state $s_0$ in the first box and calculate the safety or the reach-avoid probability of the model starting from that initial state, 
by pressing the button \boxname{Calculate}.
The button \boxname{Properties of s0} gives the abstracted state associated to $s_0$,
namely $z = \xi(\Xi(s_0))$ (cf. Algorithm \ref{algo:MC_app}), 
and all the labels assigned to this state.
\end{itemize}
\end{document}

%% file: guiwithnumbers.tikz
\begin{tikzpicture}
\node[anchor=south west,inner sep=0] (image) at (0,0) {\includegraphics[width=\linewidth]{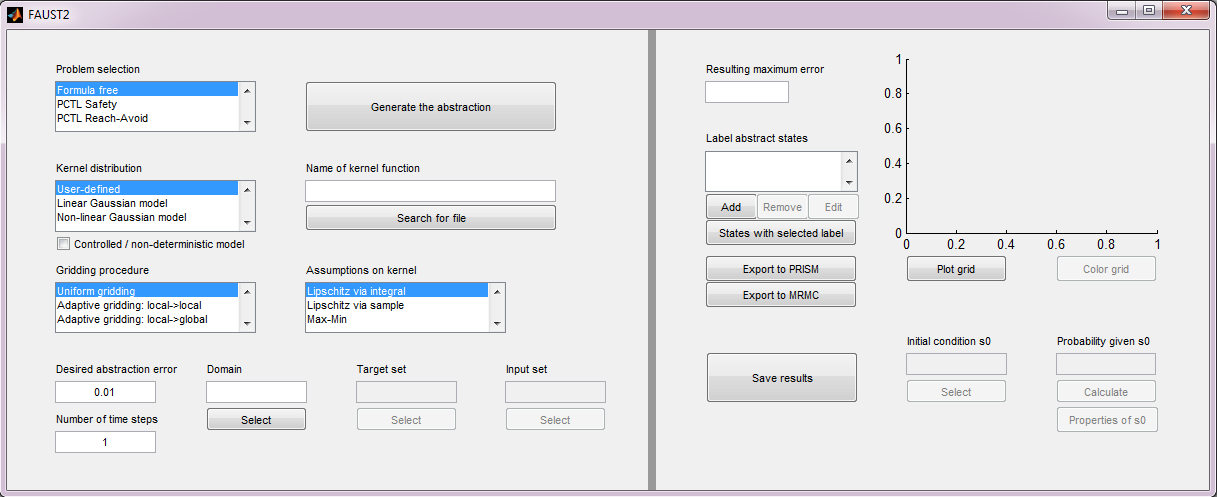}};
\begin{scope}[x={(image.south east)},y={(image.north west)}]

\draw[red,thick,rounded corners] (0.041,0.725) rectangle (0.219,0.88);
\node[red,thick,shape=circle,draw,inner sep=1pt] (test) at (0.02,0.8){1};

\draw[red,thick,rounded corners] (0.245,0.725) rectangle (0.465,0.845);
\node[red,thick,shape=circle,draw,inner sep=1pt] (test) at (0.485,0.78){6};

\draw[red,thick,rounded corners] (0.041,0.53) rectangle (0.219,0.68);
\node[red,thick,shape=circle,draw,inner sep=1pt] (test) at (0.02,0.62){2};

\draw[red,thick,rounded corners] (0.245,0.525) rectangle (0.465,0.68);
\node[red,thick,shape=circle,draw,inner sep=1pt] (test) at (0.485,0.6){7};

\draw[red,thick,rounded corners] (0.041,0.49) rectangle (0.21,0.53);
\node[red,thick,shape=circle,draw,inner sep=1pt] (test) at (0.02,0.51){3};

\draw[red,thick,rounded corners] (0.041,0.32) rectangle (0.424,0.48);
\node[red,thick,shape=circle,draw,inner sep=1pt] (test) at (0.02,0.4){4};

\draw[red,thick,rounded corners] (0.041,0.08) rectangle (0.155,0.28);
\node[red,thick,shape=circle,draw,inner sep=1pt] (test) at (0.02,0.2){5};

\draw[red,thick,rounded corners] (0.165,0.12) rectangle (0.505,0.28);
\node[red,thick,shape=circle,draw,inner sep=1pt] (test) at (0.49,0.075){8};


\draw[red,thick,rounded corners] (0.575,0.785) rectangle (0.705,0.88);
\node[red,thick,shape=circle,draw,inner sep=1pt] (test) at (0.553,0.83){9};

\draw[red,thick,rounded corners] (0.575,0.50) rectangle (0.71,0.74);
\node[red,thick,shape=circle,draw,inner sep=0.5pt] (test) at (0.553,0.62){10};

\draw[red,thick,rounded corners] (0.575,0.375) rectangle (0.71,0.495);
\node[red,thick,shape=circle,draw,inner sep=0.5pt] (test) at (0.553,0.41){11};

\draw[red,thick,rounded corners] (0.72,0.425) rectangle (0.96,0.9);
\node[red,thick,shape=circle,draw,inner sep=0.5pt] (test) at (0.982,0.7){13};

\draw[red,thick,rounded corners] (0.575,0.185) rectangle (0.71,0.305);
\node[red,thick,shape=circle,draw,inner sep=0.5pt] (test) at (0.553,0.24){12};

\draw[red,thick,rounded corners] (0.735,0.12) rectangle (0.97,0.34);
\node[red,thick,shape=circle,draw,inner sep=0.5pt] (test) at (0.975,0.07){14};

\end{scope}
\end{tikzpicture}

%% file: states_3d.tikz
\begin{tikzpicture}

\begin{axis}[%
width=3.79861111111111in,
height=3.79861111111111in,
colormap/bluered,
point meta min=9.2537236965043,
point meta max=26.5914606266758,
view={185}{15},
scale only axis,
xmin=19.5,
xmax=20.5,
xlabel={$s_1$},
ymin=19.5,
ymax=20.5,
ylabel={$s_2$},
zmin=4.5,
zmax=5.5,
zlabel={$s_3$},
axis x line*=bottom,
axis y line*=left,
axis z line*=left,
colorbar,
colorbar style={title={Prob. [\%]}},
point meta min=9.2537236965043,
point meta max=26.5914606266758
]
\addplot3[area legend,solid,line width=2.8pt,mesh,forget plot]
table[row sep=crcr, point meta=\thisrow{c}]{
x y z c\\
19.5 19.5 4.5 10.2305 \\
19.75 19.5 4.5 10.2305 \\
19.75 19.5 5 10.2305 \\
19.5 19.5 5 10.2305 \\
19.5 19.75 5 10.2305 \\
19.75 19.75 5 10.2305 \\
19.75 19.5 5 10.2305 \\
19.75 19.75 5 10.2305 \\
19.75 19.75 4.5 10.2305 \\
19.75 19.5 4.5 10.2305 \\
19.75 19.75 4.5 10.2305 \\
19.5 19.75 4.5 10.2305 \\
19.5 19.5 4.5 10.2305 \\
19.5 19.75 4.5 10.2305 \\
19.5 19.75 5 10.2305 \\
19.5 19.5 5 10.2305 \\
19.5 19.5 4.5 10.2305 \\
};
\addplot3[area legend,solid,line width=2.8pt,mesh,forget plot]
table[row sep=crcr, point meta=\thisrow{c}]{
x y z c\\
19.5 19.75 4.5 14.9573 \\
19.75 19.75 4.5 14.9573 \\
19.75 19.75 5 14.9573 \\
19.5 19.75 5 14.9573 \\
19.5 20 5 14.9573 \\
19.75 20 5 14.9573 \\
19.75 19.75 5 14.9573 \\
19.75 20 5 14.9573 \\
19.75 20 4.5 14.9573 \\
19.75 19.75 4.5 14.9573 \\
19.75 20 4.5 14.9573 \\
19.5 20 4.5 14.9573 \\
19.5 19.75 4.5 14.9573 \\
19.5 20 4.5 14.9573 \\
19.5 20 5 14.9573 \\
19.5 19.75 5 14.9573 \\
};
\addplot3[area legend,solid,line width=2.8pt,mesh,forget plot]
table[row sep=crcr, point meta=\thisrow{c}]{
x y z c\\
19.75 19.5 4.5 14.9633 \\
20 19.5 4.5 14.9633 \\
20 19.5 5 14.9633 \\
19.75 19.5 5 14.9633 \\
19.75 19.75 5 14.9633 \\
20 19.75 5 14.9633 \\
20 19.5 5 14.9633 \\
20 19.75 5 14.9633 \\
20 19.75 4.5 14.9633 \\
20 19.5 4.5 14.9633 \\
20 19.75 4.5 14.9633 \\
19.75 19.75 4.5 14.9633 \\
19.75 19.5 4.5 14.9633 \\
19.75 19.75 4.5 14.9633 \\
19.75 19.75 5 14.9633 \\
19.75 19.5 5 14.9633 \\
};
\addplot3[area legend,solid,line width=2.8pt,mesh,forget plot]
table[row sep=crcr, point meta=\thisrow{c}]{
x y z c\\
19.75 19.75 4.5 21.6732 \\
20 19.75 4.5 21.6732 \\
20 19.75 5 21.6732 \\
19.75 19.75 5 21.6732 \\
19.75 20 5 21.6732 \\
20 20 5 21.6732 \\
20 19.75 5 21.6732 \\
20 20 5 21.6732 \\
20 20 4.5 21.6732 \\
20 19.75 4.5 21.6732 \\
20 20 4.5 21.6732 \\
19.75 20 4.5 21.6732 \\
19.75 19.75 4.5 21.6732 \\
19.75 20 4.5 21.6732 \\
19.75 20 5 21.6732 \\
19.75 19.75 5 21.6732 \\
};
\addplot3[area legend,solid,line width=2.8pt,mesh,forget plot]
table[row sep=crcr, point meta=\thisrow{c}]{
x y z c\\
19.5 19.5 5 9.25372 \\
19.75 19.5 5 9.25372 \\
19.75 19.5 5.5 9.25372 \\
19.5 19.5 5.5 9.25372 \\
19.5 19.75 5.5 9.25372 \\
19.75 19.75 5.5 9.25372 \\
19.75 19.5 5.5 9.25372 \\
19.75 19.75 5.5 9.25372 \\
19.75 19.75 5 9.25372 \\
19.75 19.5 5 9.25372 \\
19.75 19.75 5 9.25372 \\
19.5 19.75 5 9.25372 \\
19.5 19.5 5 9.25372 \\
19.5 19.75 5 9.25372 \\
19.5 19.75 5.5 9.25372 \\
19.5 19.5 5.5 9.25372 \\
19.5 19.5 5 9.25372 \\
};
\addplot3[area legend,solid,line width=2.8pt,mesh,forget plot]
table[row sep=crcr, point meta=\thisrow{c}]{
x y z c\\
19.5 19.75 5 13.1184 \\
19.75 19.75 5 13.1184 \\
19.75 19.75 5.5 13.1184 \\
19.5 19.75 5.5 13.1184 \\
19.5 20 5.5 13.1184 \\
19.75 20 5.5 13.1184 \\
19.75 19.75 5.5 13.1184 \\
19.75 20 5.5 13.1184 \\
19.75 20 5 13.1184 \\
19.75 19.75 5 13.1184 \\
19.75 20 5 13.1184 \\
19.5 20 5 13.1184 \\
19.5 19.75 5 13.1184 \\
19.5 20 5 13.1184 \\
19.5 20 5.5 13.1184 \\
19.5 19.75 5.5 13.1184 \\
};
\addplot3[area legend,solid,line width=2.8pt,mesh,forget plot]
table[row sep=crcr, point meta=\thisrow{c}]{
x y z c\\
19.75 19.5 5 13.196 \\
20 19.5 5 13.196 \\
20 19.5 5.5 13.196 \\
19.75 19.5 5.5 13.196 \\
19.75 19.75 5.5 13.196 \\
20 19.75 5.5 13.196 \\
20 19.5 5.5 13.196 \\
20 19.75 5.5 13.196 \\
20 19.75 5 13.196 \\
20 19.5 5 13.196 \\
20 19.75 5 13.196 \\
19.75 19.75 5 13.196 \\
19.75 19.5 5 13.196 \\
19.75 19.75 5 13.196 \\
19.75 19.75 5.5 13.196 \\
19.75 19.5 5.5 13.196 \\
};
\addplot3[area legend,solid,line width=2.8pt,mesh,forget plot]
table[row sep=crcr, point meta=\thisrow{c}]{
x y z c\\
19.75 19.75 5 18.5534 \\
20 19.75 5 18.5534 \\
20 19.75 5.5 18.5534 \\
19.75 19.75 5.5 18.5534 \\
19.75 20 5.5 18.5534 \\
20 20 5.5 18.5534 \\
20 19.75 5.5 18.5534 \\
20 20 5.5 18.5534 \\
20 20 5 18.5534 \\
20 19.75 5 18.5534 \\
20 20 5 18.5534 \\
19.75 20 5 18.5534 \\
19.75 19.75 5 18.5534 \\
19.75 20 5 18.5534 \\
19.75 20 5.5 18.5534 \\
19.75 19.75 5.5 18.5534 \\
};
\addplot3[area legend,solid,line width=2.8pt,mesh,forget plot]
table[row sep=crcr, point meta=\thisrow{c}]{
x y z c\\
19.5 20 4.5 17.2541 \\
19.75 20 4.5 17.2541 \\
19.75 20 5 17.2541 \\
19.5 20 5 17.2541 \\
19.5 20.25 5 17.2541 \\
19.75 20.25 5 17.2541 \\
19.75 20 5 17.2541 \\
19.75 20.25 5 17.2541 \\
19.75 20.25 4.5 17.2541 \\
19.75 20 4.5 17.2541 \\
19.75 20.25 4.5 17.2541 \\
19.5 20.25 4.5 17.2541 \\
19.5 20 4.5 17.2541 \\
19.5 20.25 4.5 17.2541 \\
19.5 20.25 5 17.2541 \\
19.5 20 5 17.2541 \\
};
\addplot3[area legend,solid,line width=2.8pt,mesh,forget plot]
table[row sep=crcr, point meta=\thisrow{c}]{
x y z c\\
19.5 20.25 4.5 16.99 \\
19.75 20.25 4.5 16.99 \\
19.75 20.25 5 16.99 \\
19.5 20.25 5 16.99 \\
19.5 20.5 5 16.99 \\
19.75 20.5 5 16.99 \\
19.75 20.25 5 16.99 \\
19.75 20.5 5 16.99 \\
19.75 20.5 4.5 16.99 \\
19.75 20.25 4.5 16.99 \\
19.75 20.5 4.5 16.99 \\
19.5 20.5 4.5 16.99 \\
19.5 20.25 4.5 16.99 \\
19.5 20.5 4.5 16.99 \\
19.5 20.5 5 16.99 \\
19.5 20.25 5 16.99 \\
};
\addplot3[area legend,solid,line width=2.8pt,mesh,forget plot]
table[row sep=crcr, point meta=\thisrow{c}]{
x y z c\\
19.5 20 5 13.5006 \\
19.75 20 5 13.5006 \\
19.75 20 5.5 13.5006 \\
19.5 20 5.5 13.5006 \\
19.5 20.25 5.5 13.5006 \\
19.75 20.25 5.5 13.5006 \\
19.75 20 5.5 13.5006 \\
19.75 20.25 5.5 13.5006 \\
19.75 20.25 5 13.5006 \\
19.75 20 5 13.5006 \\
19.75 20.25 5 13.5006 \\
19.5 20.25 5 13.5006 \\
19.5 20 5 13.5006 \\
19.5 20.25 5 13.5006 \\
19.5 20.25 5.5 13.5006 \\
19.5 20 5.5 13.5006 \\
};
\addplot3[area legend,solid,line width=2.8pt,mesh,forget plot]
table[row sep=crcr, point meta=\thisrow{c}]{
x y z c\\
19.5 20.25 5 10.0671 \\
19.75 20.25 5 10.0671 \\
19.75 20.25 5.5 10.0671 \\
19.5 20.25 5.5 10.0671 \\
19.5 20.5 5.5 10.0671 \\
19.75 20.5 5.5 10.0671 \\
19.75 20.25 5.5 10.0671 \\
19.75 20.5 5.5 10.0671 \\
19.75 20.5 5 10.0671 \\
19.75 20.25 5 10.0671 \\
19.75 20.5 5 10.0671 \\
19.5 20.5 5 10.0671 \\
19.5 20.25 5 10.0671 \\
19.5 20.5 5 10.0671 \\
19.5 20.5 5.5 10.0671 \\
19.5 20.25 5.5 10.0671 \\
};
\addplot3[area legend,solid,line width=2.8pt,mesh,forget plot]
table[row sep=crcr, point meta=\thisrow{c}]{
x y z c\\
19.75 20.25 5 13.9796 \\
20 20.25 5 13.9796 \\
20 20.25 5.5 13.9796 \\
19.75 20.25 5.5 13.9796 \\
19.75 20.5 5.5 13.9796 \\
20 20.5 5.5 13.9796 \\
20 20.25 5.5 13.9796 \\
20 20.5 5.5 13.9796 \\
20 20.5 5 13.9796 \\
20 20.25 5 13.9796 \\
20 20.5 5 13.9796 \\
19.75 20.5 5 13.9796 \\
19.75 20.25 5 13.9796 \\
19.75 20.5 5 13.9796 \\
19.75 20.5 5.5 13.9796 \\
19.75 20.25 5.5 13.9796 \\
};
\addplot3[area legend,solid,line width=2.8pt,mesh,forget plot]
table[row sep=crcr, point meta=\thisrow{c}]{
x y z c\\
20 19.5 4.5 16.9503 \\
20.25 19.5 4.5 16.9503 \\
20.25 19.5 5 16.9503 \\
20 19.5 5 16.9503 \\
20 19.75 5 16.9503 \\
20.25 19.75 5 16.9503 \\
20.25 19.5 5 16.9503 \\
20.25 19.75 5 16.9503 \\
20.25 19.75 4.5 16.9503 \\
20.25 19.5 4.5 16.9503 \\
20.25 19.75 4.5 16.9503 \\
20 19.75 4.5 16.9503 \\
20 19.5 4.5 16.9503 \\
20 19.75 4.5 16.9503 \\
20 19.75 5 16.9503 \\
20 19.5 5 16.9503 \\
};
\addplot3[area legend,solid,line width=2.8pt,mesh,forget plot]
table[row sep=crcr, point meta=\thisrow{c}]{
x y z c\\
20.25 19.5 4.5 17.6758 \\
20.5 19.5 4.5 17.6758 \\
20.5 19.5 5 17.6758 \\
20.25 19.5 5 17.6758 \\
20.25 19.75 5 17.6758 \\
20.5 19.75 5 17.6758 \\
20.5 19.5 5 17.6758 \\
20.5 19.75 5 17.6758 \\
20.5 19.75 4.5 17.6758 \\
20.5 19.5 4.5 17.6758 \\
20.5 19.75 4.5 17.6758 \\
20.25 19.75 4.5 17.6758 \\
20.25 19.5 4.5 17.6758 \\
20.25 19.75 4.5 17.6758 \\
20.25 19.75 5 17.6758 \\
20.25 19.5 5 17.6758 \\
};
\addplot3[area legend,solid,line width=2.8pt,mesh,forget plot]
table[row sep=crcr, point meta=\thisrow{c}]{
x y z c\\
20 19.5 5 13.7511 \\
20.25 19.5 5 13.7511 \\
20.25 19.5 5.5 13.7511 \\
20 19.5 5.5 13.7511 \\
20 19.75 5.5 13.7511 \\
20.25 19.75 5.5 13.7511 \\
20.25 19.5 5.5 13.7511 \\
20.25 19.75 5.5 13.7511 \\
20.25 19.75 5 13.7511 \\
20.25 19.5 5 13.7511 \\
20.25 19.75 5 13.7511 \\
20 19.75 5 13.7511 \\
20 19.5 5 13.7511 \\
20 19.75 5 13.7511 \\
20 19.75 5.5 13.7511 \\
20 19.5 5.5 13.7511 \\
};
\addplot3[area legend,solid,line width=2.8pt,mesh,forget plot]
table[row sep=crcr, point meta=\thisrow{c}]{
x y z c\\
20.25 19.5 5 11.21 \\
20.5 19.5 5 11.21 \\
20.5 19.5 5.5 11.21 \\
20.25 19.5 5.5 11.21 \\
20.25 19.75 5.5 11.21 \\
20.5 19.75 5.5 11.21 \\
20.5 19.5 5.5 11.21 \\
20.5 19.75 5.5 11.21 \\
20.5 19.75 5 11.21 \\
20.5 19.5 5 11.21 \\
20.5 19.75 5 11.21 \\
20.25 19.75 5 11.21 \\
20.25 19.5 5 11.21 \\
20.25 19.75 5 11.21 \\
20.25 19.75 5.5 11.21 \\
20.25 19.5 5.5 11.21 \\
};
\addplot3[area legend,solid,line width=2.8pt,mesh,forget plot]
table[row sep=crcr, point meta=\thisrow{c}]{
x y z c\\
20 20.25 5 14.1753 \\
20.25 20.25 5 14.1753 \\
20.25 20.25 5.5 14.1753 \\
20 20.25 5.5 14.1753 \\
20 20.5 5.5 14.1753 \\
20.25 20.5 5.5 14.1753 \\
20.25 20.25 5.5 14.1753 \\
20.25 20.5 5.5 14.1753 \\
20.25 20.5 5 14.1753 \\
20.25 20.25 5 14.1753 \\
20.25 20.5 5 14.1753 \\
20 20.5 5 14.1753 \\
20 20.25 5 14.1753 \\
20 20.5 5 14.1753 \\
20 20.5 5.5 14.1753 \\
20 20.25 5.5 14.1753 \\
};
\addplot3[area legend,solid,line width=2.8pt,mesh,forget plot]
table[row sep=crcr, point meta=\thisrow{c}]{
x y z c\\
20.25 20.25 5 11.2577 \\
20.5 20.25 5 11.2577 \\
20.5 20.25 5.5 11.2577 \\
20.25 20.25 5.5 11.2577 \\
20.25 20.5 5.5 11.2577 \\
20.5 20.5 5.5 11.2577 \\
20.5 20.25 5.5 11.2577 \\
20.5 20.5 5.5 11.2577 \\
20.5 20.5 5 11.2577 \\
20.5 20.25 5 11.2577 \\
20.5 20.5 5 11.2577 \\
20.25 20.5 5 11.2577 \\
20.25 20.25 5 11.2577 \\
20.25 20.5 5 11.2577 \\
20.25 20.5 5.5 11.2577 \\
20.25 20.25 5.5 11.2577 \\
};
\addplot3[area legend,solid,line width=2.8pt,mesh,forget plot]
table[row sep=crcr, point meta=\thisrow{c}]{
x y z c\\
19.75 20 4.5 23.6366 \\
20 20 4.5 23.6366 \\
20 20 4.75 23.6366 \\
19.75 20 4.75 23.6366 \\
19.75 20.25 4.75 23.6366 \\
20 20.25 4.75 23.6366 \\
20 20 4.75 23.6366 \\
20 20.25 4.75 23.6366 \\
20 20.25 4.5 23.6366 \\
20 20 4.5 23.6366 \\
20 20.25 4.5 23.6366 \\
19.75 20.25 4.5 23.6366 \\
19.75 20 4.5 23.6366 \\
19.75 20.25 4.5 23.6366 \\
19.75 20.25 4.75 23.6366 \\
19.75 20 4.75 23.6366 \\
};
\addplot3[area legend,solid,line width=2.8pt,mesh,forget plot]
table[row sep=crcr, point meta=\thisrow{c}]{
x y z c\\
19.75 20 4.75 24.9254 \\
20 20 4.75 24.9254 \\
20 20 5 24.9254 \\
19.75 20 5 24.9254 \\
19.75 20.25 5 24.9254 \\
20 20.25 5 24.9254 \\
20 20 5 24.9254 \\
20 20.25 5 24.9254 \\
20 20.25 4.75 24.9254 \\
20 20 4.75 24.9254 \\
20 20.25 4.75 24.9254 \\
19.75 20.25 4.75 24.9254 \\
19.75 20 4.75 24.9254 \\
19.75 20.25 4.75 24.9254 \\
19.75 20.25 5 24.9254 \\
19.75 20 5 24.9254 \\
};
\addplot3[area legend,solid,line width=2.8pt,mesh,forget plot]
table[row sep=crcr, point meta=\thisrow{c}]{
x y z c\\
19.75 20.25 4.5 24.8972 \\
20 20.25 4.5 24.8972 \\
20 20.25 4.75 24.8972 \\
19.75 20.25 4.75 24.8972 \\
19.75 20.5 4.75 24.8972 \\
20 20.5 4.75 24.8972 \\
20 20.25 4.75 24.8972 \\
20 20.5 4.75 24.8972 \\
20 20.5 4.5 24.8972 \\
20 20.25 4.5 24.8972 \\
20 20.5 4.5 24.8972 \\
19.75 20.5 4.5 24.8972 \\
19.75 20.25 4.5 24.8972 \\
19.75 20.5 4.5 24.8972 \\
19.75 20.5 4.75 24.8972 \\
19.75 20.25 4.75 24.8972 \\
};
\addplot3[area legend,solid,line width=2.8pt,mesh,forget plot]
table[row sep=crcr, point meta=\thisrow{c}]{
x y z c\\
19.75 20.25 4.75 22.4311 \\
20 20.25 4.75 22.4311 \\
20 20.25 5 22.4311 \\
19.75 20.25 5 22.4311 \\
19.75 20.5 5 22.4311 \\
20 20.5 5 22.4311 \\
20 20.25 5 22.4311 \\
20 20.5 5 22.4311 \\
20 20.5 4.75 22.4311 \\
20 20.25 4.75 22.4311 \\
20 20.5 4.75 22.4311 \\
19.75 20.5 4.75 22.4311 \\
19.75 20.25 4.75 22.4311 \\
19.75 20.5 4.75 22.4311 \\
19.75 20.5 5 22.4311 \\
19.75 20.25 5 22.4311 \\
};
\addplot3[area legend,solid,line width=2.8pt,mesh,forget plot]
table[row sep=crcr, point meta=\thisrow{c}]{
x y z c\\
19.75 20 5 22.1143 \\
20 20 5 22.1143 \\
20 20 5.25 22.1143 \\
19.75 20 5.25 22.1143 \\
19.75 20.25 5.25 22.1143 \\
20 20.25 5.25 22.1143 \\
20 20 5.25 22.1143 \\
20 20.25 5.25 22.1143 \\
20 20.25 5 22.1143 \\
20 20 5 22.1143 \\
20 20.25 5 22.1143 \\
19.75 20.25 5 22.1143 \\
19.75 20 5 22.1143 \\
19.75 20.25 5 22.1143 \\
19.75 20.25 5.25 22.1143 \\
19.75 20 5.25 22.1143 \\
};
\addplot3[area legend,solid,line width=2.8pt,mesh,forget plot]
table[row sep=crcr, point meta=\thisrow{c}]{
x y z c\\
19.75 20 5.25 14.974 \\
20 20 5.25 14.974 \\
20 20 5.5 14.974 \\
19.75 20 5.5 14.974 \\
19.75 20.25 5.5 14.974 \\
20 20.25 5.5 14.974 \\
20 20 5.5 14.974 \\
20 20.25 5.5 14.974 \\
20 20.25 5.25 14.974 \\
20 20 5.25 14.974 \\
20 20.25 5.25 14.974 \\
19.75 20.25 5.25 14.974 \\
19.75 20 5.25 14.974 \\
19.75 20.25 5.25 14.974 \\
19.75 20.25 5.5 14.974 \\
19.75 20 5.5 14.974 \\
};
\addplot3[area legend,solid,line width=2.8pt,mesh,forget plot]
table[row sep=crcr, point meta=\thisrow{c}]{
x y z c\\
20 19.75 4.5 22.0667 \\
20.25 19.75 4.5 22.0667 \\
20.25 19.75 4.75 22.0667 \\
20 19.75 4.75 22.0667 \\
20 20 4.75 22.0667 \\
20.25 20 4.75 22.0667 \\
20.25 19.75 4.75 22.0667 \\
20.25 20 4.75 22.0667 \\
20.25 20 4.5 22.0667 \\
20.25 19.75 4.5 22.0667 \\
20.25 20 4.5 22.0667 \\
20 20 4.5 22.0667 \\
20 19.75 4.5 22.0667 \\
20 20 4.5 22.0667 \\
20 20 4.75 22.0667 \\
20 19.75 4.75 22.0667 \\
};
\addplot3[area legend,solid,line width=2.8pt,mesh,forget plot]
table[row sep=crcr, point meta=\thisrow{c}]{
x y z c\\
20 19.75 4.75 25.143 \\
20.25 19.75 4.75 25.143 \\
20.25 19.75 5 25.143 \\
20 19.75 5 25.143 \\
20 20 5 25.143 \\
20.25 20 5 25.143 \\
20.25 19.75 5 25.143 \\
20.25 20 5 25.143 \\
20.25 20 4.75 25.143 \\
20.25 19.75 4.75 25.143 \\
20.25 20 4.75 25.143 \\
20 20 4.75 25.143 \\
20 19.75 4.75 25.143 \\
20 20 4.75 25.143 \\
20 20 5 25.143 \\
20 19.75 5 25.143 \\
};
\addplot3[area legend,solid,line width=2.8pt,mesh,forget plot]
table[row sep=crcr, point meta=\thisrow{c}]{
x y z c\\
20.25 19.75 4.5 24.3707 \\
20.5 19.75 4.5 24.3707 \\
20.5 19.75 4.75 24.3707 \\
20.25 19.75 4.75 24.3707 \\
20.25 20 4.75 24.3707 \\
20.5 20 4.75 24.3707 \\
20.5 19.75 4.75 24.3707 \\
20.5 20 4.75 24.3707 \\
20.5 20 4.5 24.3707 \\
20.5 19.75 4.5 24.3707 \\
20.5 20 4.5 24.3707 \\
20.25 20 4.5 24.3707 \\
20.25 19.75 4.5 24.3707 \\
20.25 20 4.5 24.3707 \\
20.25 20 4.75 24.3707 \\
20.25 19.75 4.75 24.3707 \\
};
\addplot3[area legend,solid,line width=2.8pt,mesh,forget plot]
table[row sep=crcr, point meta=\thisrow{c}]{
x y z c\\
20.25 19.75 4.75 24.4095 \\
20.5 19.75 4.75 24.4095 \\
20.5 19.75 5 24.4095 \\
20.25 19.75 5 24.4095 \\
20.25 20 5 24.4095 \\
20.5 20 5 24.4095 \\
20.5 19.75 5 24.4095 \\
20.5 20 5 24.4095 \\
20.5 20 4.75 24.4095 \\
20.5 19.75 4.75 24.4095 \\
20.5 20 4.75 24.4095 \\
20.25 20 4.75 24.4095 \\
20.25 19.75 4.75 24.4095 \\
20.25 20 4.75 24.4095 \\
20.25 20 5 24.4095 \\
20.25 19.75 5 24.4095 \\
};
\addplot3[area legend,solid,line width=2.8pt,mesh,forget plot]
table[row sep=crcr, point meta=\thisrow{c}]{
x y z c\\
20 19.75 5 22.4132 \\
20.25 19.75 5 22.4132 \\
20.25 19.75 5.25 22.4132 \\
20 19.75 5.25 22.4132 \\
20 20 5.25 22.4132 \\
20.25 20 5.25 22.4132 \\
20.25 19.75 5.25 22.4132 \\
20.25 20 5.25 22.4132 \\
20.25 20 5 22.4132 \\
20.25 19.75 5 22.4132 \\
20.25 20 5 22.4132 \\
20 20 5 22.4132 \\
20 19.75 5 22.4132 \\
20 20 5 22.4132 \\
20 20 5.25 22.4132 \\
20 19.75 5.25 22.4132 \\
};
\addplot3[area legend,solid,line width=2.8pt,mesh,forget plot]
table[row sep=crcr, point meta=\thisrow{c}]{
x y z c\\
20 19.75 5.25 15.1255 \\
20.25 19.75 5.25 15.1255 \\
20.25 19.75 5.5 15.1255 \\
20 19.75 5.5 15.1255 \\
20 20 5.5 15.1255 \\
20.25 20 5.5 15.1255 \\
20.25 19.75 5.5 15.1255 \\
20.25 20 5.5 15.1255 \\
20.25 20 5.25 15.1255 \\
20.25 19.75 5.25 15.1255 \\
20.25 20 5.25 15.1255 \\
20 20 5.25 15.1255 \\
20 19.75 5.25 15.1255 \\
20 20 5.25 15.1255 \\
20 20 5.5 15.1255 \\
20 19.75 5.5 15.1255 \\
};
\addplot3[area legend,solid,line width=2.8pt,mesh,forget plot]
table[row sep=crcr, point meta=\thisrow{c}]{
x y z c\\
20.25 19.75 5 19.2269 \\
20.5 19.75 5 19.2269 \\
20.5 19.75 5.25 19.2269 \\
20.25 19.75 5.25 19.2269 \\
20.25 20 5.25 19.2269 \\
20.5 20 5.25 19.2269 \\
20.5 19.75 5.25 19.2269 \\
20.5 20 5.25 19.2269 \\
20.5 20 5 19.2269 \\
20.5 19.75 5 19.2269 \\
20.5 20 5 19.2269 \\
20.25 20 5 19.2269 \\
20.25 19.75 5 19.2269 \\
20.25 20 5 19.2269 \\
20.25 20 5.25 19.2269 \\
20.25 19.75 5.25 19.2269 \\
};
\addplot3[area legend,solid,line width=2.8pt,mesh,forget plot]
table[row sep=crcr, point meta=\thisrow{c}]{
x y z c\\
20.25 19.75 5.25 11.5628 \\
20.5 19.75 5.25 11.5628 \\
20.5 19.75 5.5 11.5628 \\
20.25 19.75 5.5 11.5628 \\
20.25 20 5.5 11.5628 \\
20.5 20 5.5 11.5628 \\
20.5 19.75 5.5 11.5628 \\
20.5 20 5.5 11.5628 \\
20.5 20 5.25 11.5628 \\
20.5 19.75 5.25 11.5628 \\
20.5 20 5.25 11.5628 \\
20.25 20 5.25 11.5628 \\
20.25 19.75 5.25 11.5628 \\
20.25 20 5.25 11.5628 \\
20.25 20 5.5 11.5628 \\
20.25 19.75 5.5 11.5628 \\
};
\addplot3[area legend,solid,line width=2.8pt,mesh,forget plot]
table[row sep=crcr, point meta=\thisrow{c}]{
x y z c\\
20 20 4.5 25.9497 \\
20.25 20 4.5 25.9497 \\
20.25 20 4.75 25.9497 \\
20 20 4.75 25.9497 \\
20 20.25 4.75 25.9497 \\
20.25 20.25 4.75 25.9497 \\
20.25 20 4.75 25.9497 \\
20.25 20.25 4.75 25.9497 \\
20.25 20.25 4.5 25.9497 \\
20.25 20 4.5 25.9497 \\
20.25 20.25 4.5 25.9497 \\
20 20.25 4.5 25.9497 \\
20 20 4.5 25.9497 \\
20 20.25 4.5 25.9497 \\
20 20.25 4.75 25.9497 \\
20 20 4.75 25.9497 \\
};
\addplot3[area legend,solid,line width=2.8pt,mesh,forget plot]
table[row sep=crcr, point meta=\thisrow{c}]{
x y z c\\
20 20 4.75 26.3839 \\
20.25 20 4.75 26.3839 \\
20.25 20 5 26.3839 \\
20 20 5 26.3839 \\
20 20.25 5 26.3839 \\
20.25 20.25 5 26.3839 \\
20.25 20 5 26.3839 \\
20.25 20.25 5 26.3839 \\
20.25 20.25 4.75 26.3839 \\
20.25 20 4.75 26.3839 \\
20.25 20.25 4.75 26.3839 \\
20 20.25 4.75 26.3839 \\
20 20 4.75 26.3839 \\
20 20.25 4.75 26.3839 \\
20 20.25 5 26.3839 \\
20 20 5 26.3839 \\
};
\addplot3[area legend,solid,line width=2.8pt,mesh,forget plot]
table[row sep=crcr, point meta=\thisrow{c}]{
x y z c\\
20 20.25 4.5 26.3602 \\
20.25 20.25 4.5 26.3602 \\
20.25 20.25 4.75 26.3602 \\
20 20.25 4.75 26.3602 \\
20 20.5 4.75 26.3602 \\
20.25 20.5 4.75 26.3602 \\
20.25 20.25 4.75 26.3602 \\
20.25 20.5 4.75 26.3602 \\
20.25 20.5 4.5 26.3602 \\
20.25 20.25 4.5 26.3602 \\
20.25 20.5 4.5 26.3602 \\
20 20.5 4.5 26.3602 \\
20 20.25 4.5 26.3602 \\
20 20.5 4.5 26.3602 \\
20 20.5 4.75 26.3602 \\
20 20.25 4.75 26.3602 \\
};
\addplot3[area legend,solid,line width=2.8pt,mesh,forget plot]
table[row sep=crcr, point meta=\thisrow{c}]{
x y z c\\
20 20.25 4.75 23.2273 \\
20.25 20.25 4.75 23.2273 \\
20.25 20.25 5 23.2273 \\
20 20.25 5 23.2273 \\
20 20.5 5 23.2273 \\
20.25 20.5 5 23.2273 \\
20.25 20.25 5 23.2273 \\
20.25 20.5 5 23.2273 \\
20.25 20.5 4.75 23.2273 \\
20.25 20.25 4.75 23.2273 \\
20.25 20.5 4.75 23.2273 \\
20 20.5 4.75 23.2273 \\
20 20.25 4.75 23.2273 \\
20 20.5 4.75 23.2273 \\
20 20.5 5 23.2273 \\
20 20.25 5 23.2273 \\
};
\addplot3[area legend,solid,line width=2.8pt,mesh,forget plot]
table[row sep=crcr, point meta=\thisrow{c}]{
x y z c\\
20.25 20 4.5 26.5915 \\
20.5 20 4.5 26.5915 \\
20.5 20 4.75 26.5915 \\
20.25 20 4.75 26.5915 \\
20.25 20.25 4.75 26.5915 \\
20.5 20.25 4.75 26.5915 \\
20.5 20 4.75 26.5915 \\
20.5 20.25 4.75 26.5915 \\
20.5 20.25 4.5 26.5915 \\
20.5 20 4.5 26.5915 \\
20.5 20.25 4.5 26.5915 \\
20.25 20.25 4.5 26.5915 \\
20.25 20 4.5 26.5915 \\
20.25 20.25 4.5 26.5915 \\
20.25 20.25 4.75 26.5915 \\
20.25 20 4.75 26.5915 \\
};
\addplot3[area legend,solid,line width=2.8pt,mesh,forget plot]
table[row sep=crcr, point meta=\thisrow{c}]{
x y z c\\
20.25 20 4.75 25.1175 \\
20.5 20 4.75 25.1175 \\
20.5 20 5 25.1175 \\
20.25 20 5 25.1175 \\
20.25 20.25 5 25.1175 \\
20.5 20.25 5 25.1175 \\
20.5 20 5 25.1175 \\
20.5 20.25 5 25.1175 \\
20.5 20.25 4.75 25.1175 \\
20.5 20 4.75 25.1175 \\
20.5 20.25 4.75 25.1175 \\
20.25 20.25 4.75 25.1175 \\
20.25 20 4.75 25.1175 \\
20.25 20.25 4.75 25.1175 \\
20.25 20.25 5 25.1175 \\
20.25 20 5 25.1175 \\
};
\addplot3[area legend,solid,line width=2.8pt,mesh,forget plot]
table[row sep=crcr, point meta=\thisrow{c}]{
x y z c\\
20.25 20.25 4.5 25.0866 \\
20.5 20.25 4.5 25.0866 \\
20.5 20.25 4.75 25.0866 \\
20.25 20.25 4.75 25.0866 \\
20.25 20.5 4.75 25.0866 \\
20.5 20.5 4.75 25.0866 \\
20.5 20.25 4.75 25.0866 \\
20.5 20.5 4.75 25.0866 \\
20.5 20.5 4.5 25.0866 \\
20.5 20.25 4.5 25.0866 \\
20.5 20.5 4.5 25.0866 \\
20.25 20.5 4.5 25.0866 \\
20.25 20.25 4.5 25.0866 \\
20.25 20.5 4.5 25.0866 \\
20.25 20.5 4.75 25.0866 \\
20.25 20.25 4.75 25.0866 \\
};
\addplot3[area legend,solid,line width=2.8pt,mesh,forget plot]
table[row sep=crcr, point meta=\thisrow{c}]{
x y z c\\
20.25 20.25 4.75 20.3791 \\
20.5 20.25 4.75 20.3791 \\
20.5 20.25 5 20.3791 \\
20.25 20.25 5 20.3791 \\
20.25 20.5 5 20.3791 \\
20.5 20.5 5 20.3791 \\
20.5 20.25 5 20.3791 \\
20.5 20.5 5 20.3791 \\
20.5 20.5 4.75 20.3791 \\
20.5 20.25 4.75 20.3791 \\
20.5 20.5 4.75 20.3791 \\
20.25 20.5 4.75 20.3791 \\
20.25 20.25 4.75 20.3791 \\
20.25 20.5 4.75 20.3791 \\
20.25 20.5 5 20.3791 \\
20.25 20.25 5 20.3791 \\
};
\addplot3[area legend,solid,line width=2.8pt,mesh,forget plot]
table[row sep=crcr, point meta=\thisrow{c}]{
x y z c\\
20 20 5 22.8865 \\
20.25 20 5 22.8865 \\
20.25 20 5.25 22.8865 \\
20 20 5.25 22.8865 \\
20 20.25 5.25 22.8865 \\
20.25 20.25 5.25 22.8865 \\
20.25 20 5.25 22.8865 \\
20.25 20.25 5.25 22.8865 \\
20.25 20.25 5 22.8865 \\
20.25 20 5 22.8865 \\
20.25 20.25 5 22.8865 \\
20 20.25 5 22.8865 \\
20 20 5 22.8865 \\
20 20.25 5 22.8865 \\
20 20.25 5.25 22.8865 \\
20 20 5.25 22.8865 \\
};
\addplot3[area legend,solid,line width=2.8pt,mesh,forget plot]
table[row sep=crcr, point meta=\thisrow{c}]{
x y z c\\
20 20 5.25 15.1912 \\
20.25 20 5.25 15.1912 \\
20.25 20 5.5 15.1912 \\
20 20 5.5 15.1912 \\
20 20.25 5.5 15.1912 \\
20.25 20.25 5.5 15.1912 \\
20.25 20 5.5 15.1912 \\
20.25 20.25 5.5 15.1912 \\
20.25 20.25 5.25 15.1912 \\
20.25 20 5.25 15.1912 \\
20.25 20.25 5.25 15.1912 \\
20 20.25 5.25 15.1912 \\
20 20 5.25 15.1912 \\
20 20.25 5.25 15.1912 \\
20 20.25 5.5 15.1912 \\
20 20 5.5 15.1912 \\
};
\addplot3[area legend,solid,line width=2.8pt,mesh,forget plot]
table[row sep=crcr, point meta=\thisrow{c}]{
x y z c\\
20.25 20 5 19.4221 \\
20.5 20 5 19.4221 \\
20.5 20 5.25 19.4221 \\
20.25 20 5.25 19.4221 \\
20.25 20.25 5.25 19.4221 \\
20.5 20.25 5.25 19.4221 \\
20.5 20 5.25 19.4221 \\
20.5 20.25 5.25 19.4221 \\
20.5 20.25 5 19.4221 \\
20.5 20 5 19.4221 \\
20.5 20.25 5 19.4221 \\
20.25 20.25 5 19.4221 \\
20.25 20 5 19.4221 \\
20.25 20.25 5 19.4221 \\
20.25 20.25 5.25 19.4221 \\
20.25 20 5.25 19.4221 \\
};
\addplot3[area legend,solid,line width=2.8pt,mesh,forget plot]
table[row sep=crcr, point meta=\thisrow{c}]{
x y z c\\
20.25 20 5.25 11.5289 \\
20.5 20 5.25 11.5289 \\
20.5 20 5.5 11.5289 \\
20.25 20 5.5 11.5289 \\
20.25 20.25 5.5 11.5289 \\
20.5 20.25 5.5 11.5289 \\
20.5 20 5.5 11.5289 \\
20.5 20.25 5.5 11.5289 \\
20.5 20.25 5.25 11.5289 \\
20.5 20 5.25 11.5289 \\
20.5 20.25 5.25 11.5289 \\
20.25 20.25 5.25 11.5289 \\
20.25 20 5.25 11.5289 \\
20.25 20.25 5.25 11.5289 \\
20.25 20.25 5.5 11.5289 \\
20.25 20 5.5 11.5289 \\
};

\end{axis}
\end{tikzpicture}%